# LaCrO$_3$ heteroepitaxy on SrTiO$_3$(001) by molecular beam epitaxy

L. Qiao[1], T.C. Droubay[1], M.E. Bowden[2], V. Shutthanandan[2], T.C. Kaspar[1],

S. A. Chambers[1]

[1]*Fundamental and Computational Science Directorate, Pacific Northwest National Laboratory, Richland, WA 99352*

[2]*Environmental and Molecular Sciences Laboratory, Pacific Northwest National Laboratory, Richland, WA 99352*

## Abstract

Stoichiometric, epitaxial LaCrO$_3$ films have been grown on TiO$_2$-terminated SrTiO$_3$(001) substrates by molecular beam epitaxy using O$_2$ as the oxidant. Film growth occurred in a layer-by-layer fashion, giving rise to structurally excellent films and surfaces which preserve the step-terrace structure of the substrate. The critical thickness is in excess of 500 Å. Near-surface Cr(III) is highly susceptible to further oxidation to Cr(V), leading to the formation of a disordered phase upon exposure to atomic oxygen. Recovery of the original epitaxial LaCrO$_3$ phase is readily achieved by vacuum annealing.

Complex oxides exhibit a range of physical properties unparalleled by any other class of materials. Their electronic, magnetic, and optical characteristics span the field of possibilities, and a wealth of applications awaits our ability to understand and control these properties. Materials synthesis is key to realizing the full potential of complex oxides, particularly in thin-film form. Gaining understanding of and control over composition and defect density are of major importance. Moreover, different synthetic methods appear to have different effects on defect creation. For instance, record high electron mobilities in La-doped SrTiO$_3$ (STO) grown by hybrid molecular beam epitaxy (MBE) were recently demonstrated, surpassing what had been achieved previously by bulk synthesis and thin-film growth.[1] Perfecting stoichiometric control and minimizing defect creation appear to have been critical.[2]

LaCrO$_3$ (LCO) is an antiferromagnetic insulator with a charge transfer gap of 3.3 eV.[3] In the bulk, this perovskite exhibits an orthorhombic structure (space group *Pbnm*) with lattice parameters $a = 5.513$ Å, $b = 5.476$ Å, $c = 7.759$ Å.[4] However, isolation of a pseudocubic cell within the orthorhombic structure yields $a = b = c = 3.885$ Å (see structural models in Fig. 1). Thus, LCO is a candidate for heteroepitaxy on SrTiO$_3$(001), for which $a = b = c = 3.905$ Å.

In this Letter, we describe the epitaxial growth of LCO on SrTiO$_3$(001) by MBE.[5] Atomically flat, TiO$_2$-terminated SrTiO$_3$ (STO) substrates were prepared as described previously.[6] La and Cr were evaporated from an effusion cell and an electron beam evaporator, respectively, and the O$_2$ partial pressure was kept constant at ~6 × 10$^{-8}$ Torr. The La flux was measured before and after deposition by a quartz crystal oscillator (QCO), and the Cr flux was monitored and controlled during deposition by atomic absorption. The substrate temperature during growth was $650° ± 50°C$, as measured by a two-color infrared pyrometer. The $O_2$ valve was closed immediately after growth and samples were cooled to room temperature *in vacuo*. Reflection high-energy electron diffraction (RHEED) intensity oscillations and patterns were used to monitor the overall growth rate and surface structure, respectively. Film stoichiometry and oxidation states were determined by Rutherford backscattering spectrometry (RBS) and *in-situ* x-ray photoelectron spectroscopy (XPS), respectively. Lattice parameters and crystallographic quality were determined by x-ray diffraction (XRD), and surface morphologies were measured by tapping-mode atomic force microscopy (AFM).

We show in Fig. 1 RHEED specular beam intensity oscillations as a function of growth time for a 25 unit cell (u.c., referenced to the pseudocubic cell) film. Similar persistent oscillations were observed for film thicknesses up to 130 u.c., demonstrating persistent layer-by-layer growth. A typical pair of RHEED patterns for the starting substrate and the film after cooling to ambient temperature are shown as insets, along with an AFM image and a line profile. The surface periodicity evolves from (1×1) for the substrate to (2×2) for the film surface. Moreover, the film surface exhibits a well-defined terrace-step structure with typical step heights of ~ 4 Å, the $c$ value for pseudocubic LCO. This result confirms the layer-by-layer growth mode revealed by RHEED oscillations and the presence of a single surface termination, which we expect is $CrO_2$, based on the fact that all growths were terminated after an integral number of oscillations were recorded.

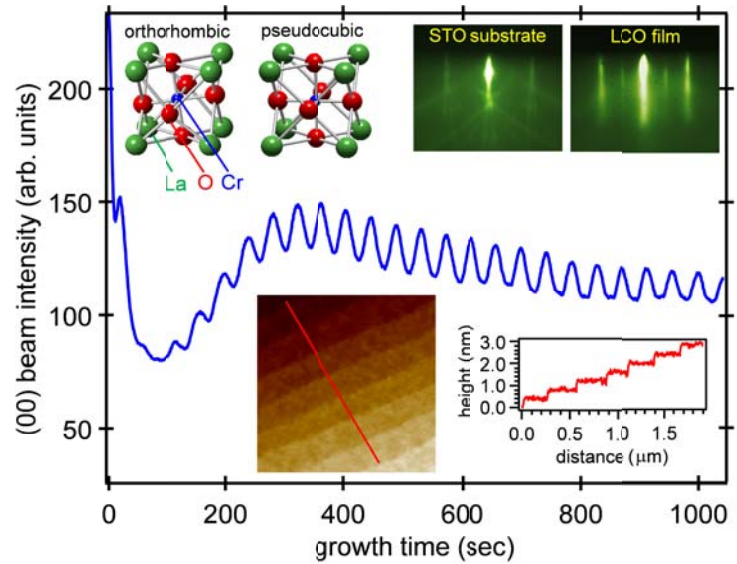

FIG. 1 RHEED intensity oscillations as a function of growth time for a 25 u.c. LCO film. Insets include structural models of orthorhombic and pseudocubic LCO, typical RHEED patterns for an LCO film and a starting STO substrate, and a typical AFM image and line scan for a LCO film.

We show in Fig. 2 a typical RBS spectrum obtained for a 25 u.c. film with the incident beam oriented 7° off normal in a random geometry and a scattering angle of 150°. Also, shown is a simulated SIMNRA[7] spectrum in which the La:Cr atom ratio is 1:1. The statistical uncertainty for the La peak area ($\sqrt{N}/N$, where $N$ is the total counts in the peak above background) is quite low (0.4%) because of the high intensity of this feature and the low background.

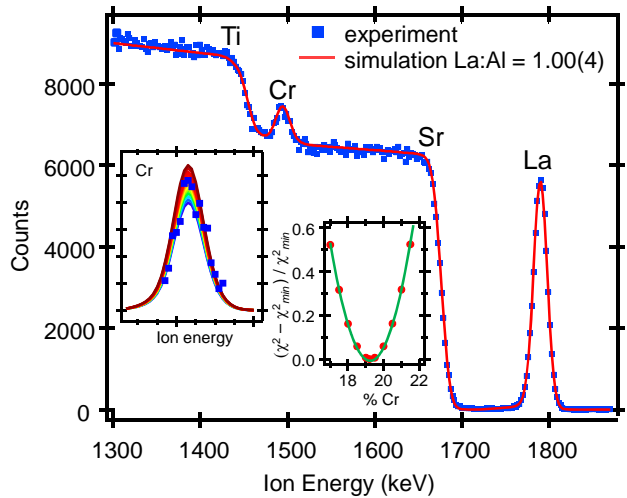

FIG. 2 Non-aligned RBS for a 25 u.c. LCO film using a 2 MeV He ion beam with the incident beam oriented 7° off normal and a scattering angle of 150°. Insets – the Cr peak along (squares) along with a family of simulations for Cr percentages ranging from 17 to 23 at. % (left), and a least-squares fit of these simulations to the Cr feature (right).

In contrast, the uncertainty in the Cr peak area is larger because the Cr peak sits on the background of the Sr plateau, and is of lower intensity due to the smaller atomic number of Cr compared to La. After subtraction of the background Sr plateau, the statistical uncertainty in the Cr peak area is ~5%. Least-squares fits of these two peaks to model SIMNRA simulations yield absolute La and Cr atomic percentages of 19.3 ± 0.1% and 19.2 ± 0.8%, respectively, leading to a La:Cr ratio of 1.00 ± 0.04. This analysis is shown for the Cr peak as insets in Fig. 2. The family of simulations in the left inset span an absolute Cr concentration from 17 to 23 at. %. The $\chi^2$ values referenced to $\chi^2$ at the minimum are plotted in the right inset. The Cr percentage range associated with $(\chi^2 - \chi^2_{min}) / \chi^2_{min} = 0.05$ below the minimum to $(\chi^2 - \chi^2_{min}) / \chi^2_{min} = 0.05$ above the minimum is our estimated uncertainty. RBS data collected on a representative set of thick LCO films were utilized to calibrate XPS peak area ratios, which were measured *in situ* on all films.

A first indication of the crystallographic quality in these films is revealed by the persistent Kiessig fringes surrounding the LCO (002) Bragg reflections in the XRD $2\theta$-$\omega$ scan, as seen in Fig. 3. Matching experiment to simulations yields a film thickness of ~ 500 Å, in good agreement with the values obtained from RHEED oscillations and RBS analysis. The film and substrate reflections are easily distinguished, and the Bragg peak angle for the film indicates a smaller *c* dimension than that for the substrate. By simultaneously fitting profiles to the (202), (113), (114), and (123) in-plane reflections (not shown), we determine $a = b = 3.904$ Å and $c = 3.881$ Å, the latter dimension fully consistent with the peak position observed in the out-of-plane scan. The film is thus fully coherent with the substrate, and the critical thickness ($h_c$) exceeds 500 Å. Based on elastic theory, $h_c$ is predicted by the Matthews-Blakeslee [8,9] and People-Bean [10,11] models to be ~ 83 and 181 Å, respectively. The much larger experimental value of $h_c$ may be due at least in part to the presence of cation intermixing at the interface, which grades the strain.[6,12] A detailed investigation of interface composition based on high-resolution RBS and scanning transmission electron microscopy/electron energy loss spectroscopy is in progress. Further evidence of film quality was seen in the XRD $\omega$ (rocking curve) scans of the (002) reflections (inset in Fig. 3). The film and substrate peaks had the same full width at half maximum (FWHM), and both sets of planes were parallel to within ±0.002°. A wide range of FWHM was measured (0.008° to 0.030°), indicating that the crystallographic quality of the substrate was the limiting factor in film quality.

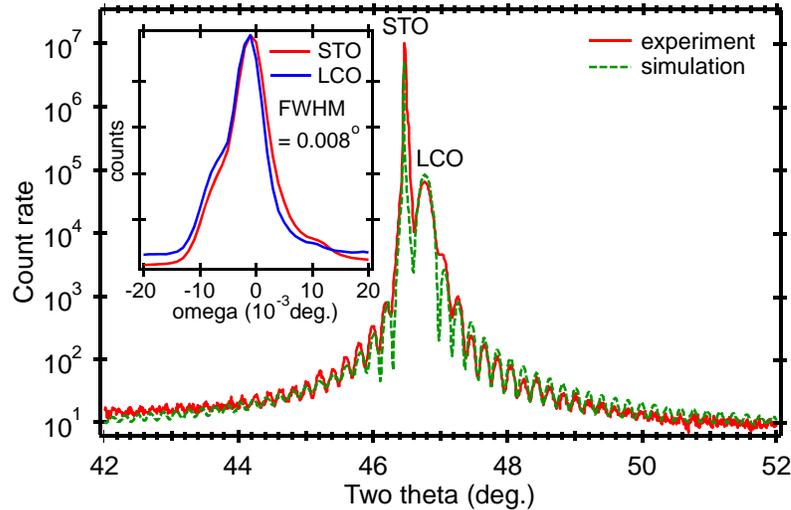

FIG. 3 Measured and simulated XRD $2\theta$-$\omega$ scan for the (002) reflection for 500 Å LCO/STO(001). Inset – (002) rocking curves for the same 500 Å LCO film.

We have found that successful growth of epitaxial $LaCrO_3$ requires the use of molecular oxygen rather than atomic oxygen from a plasma source or from ozone. Exposure of the film surface to atomic oxygen from an electron cyclotron resonance oxygen plasma at $2.0\times10^{-5}$ Torr for a few minutes causes the surface to become disordered as Cr(III) ions in the near-surface region are oxidized to a higher charge state. This result is illustrated in Fig. 4. A significant change in the Cr $2p$ XPS line shape occurs as new features appear at ~576 eV and ~586 eV in the $2p_{3/2}$ and $2p_{1/2}$ spectral regions, respectively. These new peaks are characteristic of Cr(V), as measured in polycrystalline $LaCrO_4$.[13,14] This Cr(V)-containing layer, which might be disordered $LaCrO_4$, is ~1.5 nm thick based on angle-resolved XPS and assuming uniform and complete oxidation, and is stable in high vacuum over a period of at least two weeks. Moreover, the surface can be reduced back to the original epitaxial $LaCrO_3$ phase by annealing in vacuum at the film growth temperature, as also seen in Fig 4.

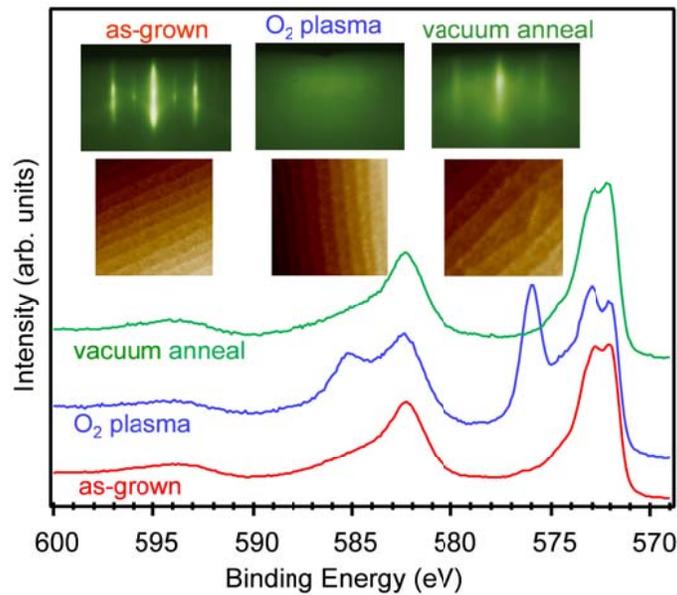

FIG. 4 Cr $2p$ XPS spectra for an as-grown (red), oxygen plasma treated (blue), and vacuum annealed (green) LCO film. Insets show the corresponding RHEED and AFM images for the film during different stages of this redox process.

The hypothesis that this surface phase is disordered $LaCrO_4$ is reasonable because $LaCrO_4$ exhibits a monoclinic structure incommensurate with STO and LCO.[15] Interestingly, this redox process at least partially preserves the terrace-step structure of the original film surface, as seen in the AFM images shown in Fig. 4. A blunting of the steps and a roughening of the terraces is evident. Additionally, the RHEED pattern exhibits higher background after recovery of the original phase, indicating a higher point defect density than was present on the original surface.

In summary, high-quality epitaxial LaCrO$_3$ has been successfully grown on TiO$_2$-terminated SrTiO$_3$(001) by MBE. Layer-by-layer growth was confirmed by both RHEED oscillations and AFM images of the final surface. Structural Cr(III) in the near-surface region is readily oxidized to a disordered phase containing Cr(V) upon exposure to atomic oxygen, and annealing *in vacuo* results in conversion back to crystalline LaCrO$_3$. This interesting result may render this complex oxide useful in fundamental studies of surface photophysics and photochemistry.

This work was supported by the Office of Science, Division of Materials Sciences and Engineering and Division of Chemical Sciences, U.S. Department of Energy, and was performed in the Environmental Molecular Sciences Laboratory, a national scientific user facility sponsored by the Office of Biological and Environmental Research of the Department of Energy and located at Pacific Northwest National Laboratory.

# References

1 J. Son, P. Moetakef, B. Jalan, O. Bierwagen, N. J. Wright, R. Engel-Herbert, S. Stemmer, Nat. Mat. **9**, 482 (2010).

2 B. Jalan, R. Engel-Herbert, N. J. Wright, and S. Stemmer, J. Vac. Sci. Technol. A **27**, 461 (2009).

3 T. Arima, Y. Tokura, and J. B. Torrance, Phys. Rev. B **48**, 17006 (1993).

4 J. Yang, Acta Crystallogr. Sect. B **64**, 281 (2008).

5 S. A. Chambers, Adv. Mater. **22**, 219 (2010).

6 L. Qiao, T. C. Droubay, T. Varga, M. E. Bowden, V. Shutthanandan, Z. Zhu, T. C. Kaspar, and S. A. Chambers, Phys. Rev. B **83**, 085408 (2011).

7 M. Mayer, (Max-Planck-Institut für Plasmaphysik, Garching, Germany, 2008), p. SIMNRA User's Guide 6.04.

8 S. Venkatesan, A. Vlooswijk, B. J. Kooi, A. Morelli, G. Palasantzas, J. T. M. De Hosson, and B. Noheda, Phys. Rev. B **78**, 104112 (2008).

9 S. M. Hu, J. Appl. Phys. **69**, 7901 (1991).

10 R. People and J. C. Bean, Appl. Phys. Lett. **47**, 322 (1985).

11 R. People and J. C. Bean, Appl. Phys. Lett. **49**, 229 (1986).

12 S. A. Chambers, M. H. Engelhard, V. Shutthanandan, Z. Zhu, T. C. Droubay, L. Qiao, P. V. Sushko, T. Feng, H. D. Lee, T. Gustafsson, E. Garfunkel, A. B. Shah, J. M. Zuo, and Q. M. Ramasse, Surf. Sci. Rep. **65**, 317 (2010).

13 H. Konno, H. Tachikawa, A. Furusaki, and R. Furuichi, Anal. Sci. **8**, 641 (1992).

14 Y. Aoki, H. Konno, H. Tachikawa, and M. Inagaki, Bull. Chem. Soc. Jpn. **73**, 1197 (2000).

[15] S. G. Manca and E. J. Baran, J. Appl. Crystallogr. **15**, 102 (1982)